# Faster-than-Nyquist Nonlinear Frequency Division Multiplexing System

Xulun Zhang[1], Lixia Xi[1*], Peng Sun[1], Zibo Zheng[1], Jiacheng Wei[1], Yue Wu[1], Shucheng Du[2], Xiaoguang Zhang[1]

[1]*State Key Laboratory of Information Photonics and Optical Communications, Beijing University of Posts and Telecommunications, Beijing, China*
[2]*School of nuclear science & technology, Beijing Normal University, Beijing, China*
xilixia@bupt.edu.cn

**Abstract:** The faster-than-Nyquist nonlinear frequency division multiplexing system was first proposed, which provided 12.5% and 25% increase in spectral efficiency for the compression factor of 0.89 and 0.8 respectively. © 2020 The Author(s)

## 1. Introduction

Nonlinear frequency division multiplexing (NFDM) has attracted great attention due to its natural resistance to Kerr nonlinearity [1-5]. Unlike the Nyquist system (time domain modulation) and OFDM (frequency domain modulation), in NFDM system, the information is modulated on the spectra of nonlinear Fourier domain (NFD), which are defined by nonlinear Fourier transform (NFT). These nonlinear spectra propagate linearly over the fiber channel, so they are used to modulate information to overcome the limit of nonlinearity[3]. However, due to both the imperfection of the NFT algorithms and the link conditions, the spectral efficiency (SE) of the NFDM system is limited. The achieved data rate and SE of demonstrated NFDM systems are still far below those of conventional linear systems [3, 6].

Faster-than-Nyquist (FTN) is one of the effective techniques to increase SE, significantly[7]. For FTN-OFDM, the subcarrier spacing in frequency domain is reduced and its SE is increased compared to OFDM, but at the expense of inter carrier interference (ICI)[8]. NFDM is analogous to OFDM, where the data is modulated independently over separate sub-carriers in NFD. In order to further increase the SE of NFDM, we extended the concept of FTN-OFDM. To the best of our knowledge, we first proposed the FTN-NFDM. The upper bound of SE for the FTN-NFDM system was derived and the FTN-NFDM transmission was demonstrated for the first time.

## 2. The spectral efficiency of NFDM system

In NFDM system, the data can be modulated on the scattering coefficient $b(\lambda)$ in NFD, which is expressed as[9]

$$b_m(\lambda) = A \sum_{k=-(N/2)}^{N/2-1} c_{m,k} \frac{\sin(\lambda T_0/T_S + k\pi)}{\lambda T_0/T_S + k\pi} e^{-2jm\lambda T_1/T_S}, \quad \lambda \in R \tag{1}$$

Where A controls the launch power, $m$ denotes the $m^{th}$ data block, $N$ is the number of nonlinear subcarriers and $C_{m,k}$ presents the QAM symbol. $T_0$ and $T_1$ are the useful burst and total block time duration (including guard interval (GI)), respectively. $T_S$ is the normalization time parameter. In order to avoid adjacent bursts interference caused by the dispersion of fiber channel, the zero GI is added and it can be estimated as $T_{GI}=2\pi B\beta_2 L$, where $\beta_2$ is the chromatic dispersion coefficient and $L$ is the transmission distance. By applying the pre-dispersion compensation (PDC), the GI can be reduced by a factor of 2[10]. Similar to OFDM, the useful burst duration $T_0$ and the number of nonlinear subcarriers $N_{NFDM}$ satisfy $T_0=N_{NFDM}/B$ for NFDM system, where $B$ is the bandwidth of signal. The upper bound of the normalized SE, in symbol/s/Hz, of NFDM system with and without PDC can be estimated as[11]

$$SE = \frac{N_{NFDM}}{(T_0+T_{GI})B} = \frac{1}{1+2\pi B^2\beta_2 L/N_{NFDM}}, \quad SE_{pre} = \frac{N_{NFDM}}{(T_0+T_{GI}/2)B} = \frac{1}{1+\pi B^2\beta_2 L/N_{NFDM}} \tag{2}$$

The normalized SE can be increased by using the PDC and more subcarriers. However, the upper bound of the normalized SE cannot exceed 1 in theory, even if a large number of nonlinear subcarriers are used.

## 3. FTN-NFDM system

FTN is an effective technique to increase SE. First, the basic knowledge of FTN-OFDM was reviewed. The frequency spectrum of an FTN-OFDM symbol can be described as

$$F(f) = A \sum_{k=-(N/2)}^{N/2-1} c_k \frac{\sin(fT\pi + k\alpha\pi)}{fT\pi + k\alpha\pi}, \quad \alpha < 1 \tag{3}$$

Where $\alpha$ is the bandwidth compression factor, $T$ is the period of FTN-OFDM symbol and $c_k$ is the QAM symbol. The bandwidth of FTN OFDM signal is compressed to $B\times\alpha$, where B is the bandwidth of OFDM($\alpha$=1) [12].

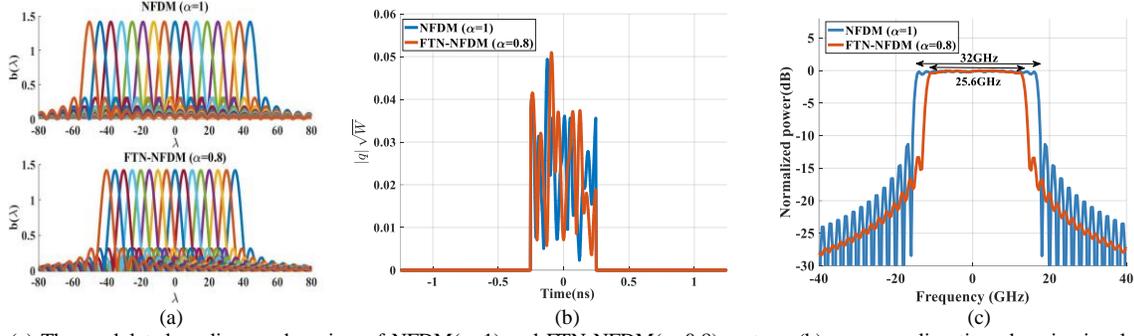

Fig.1 (a) The modulated nonlinear subcarriers of NFDM($\alpha$=1) and FTN-NFDM($\alpha$=0.8) system, (b) corresponding time domain signal, and (c) normalized power spectrum, where the number of subcarriers is 16.

In order to increase the SE of NFDM, the FTN-NFDM is proposed and the concept of FTN-OFDM is extended from frequency domain to NFD. Similarly, the FTN-NFDM signal can be expressed as

$$b_m(\lambda) = A \sum_{k=-(N/2)}^{N/2-1} c_{m,k} \frac{\sin(\lambda T_0/T_S + k\alpha\pi)}{\lambda T_0/T_S + k\alpha\pi} e^{-2jm\lambda T_1/T_S} \quad (4)$$

the bandwidth of FTN-NFDM signal also can be estimated as B× $\alpha$. An example is shown in Fig.1 (a-c), which includes the modulated nonlinear subcarriers, corresponding time domain signal and normalized power spectrum of NFDM and FTN-NFDM system, respectively.

For a given bandwidth range and useful burst time, the normalized SE of FTN NFDM with PDC can be estimated as

$$SE_{FTN-pre} = \frac{N_{NFDM}/\alpha}{(T_0+T_{GI}/2)B} = \frac{N_{FTN-NFDM}}{(T_0+T_{GI}/2)B} = \frac{1}{\alpha+\pi B^2 \beta_2 L/N_{FTN-NFDM}} = \frac{1}{\alpha} SE_{pre} \quad (5)$$

Where $N_{FTN-NFDM}$ is the number of nonlinear subcarriers in FTN-NFDM system. As shown in Fig.2 (a), the normalized SE of FTN-NFDM system is higher than that in NFDM system. The smaller the compression factor $\alpha$, the higher the normalized SE. The upper bound of FTN-NFDM system is $1/\alpha$ in theory. Like FTN-OFDM, for NFDM system, the SE was increased at the expense of ICI. The demodulated QAM symbols interfered by ICI is shown in Fig.2 (b, c), where the impact of phase noises were not considered.

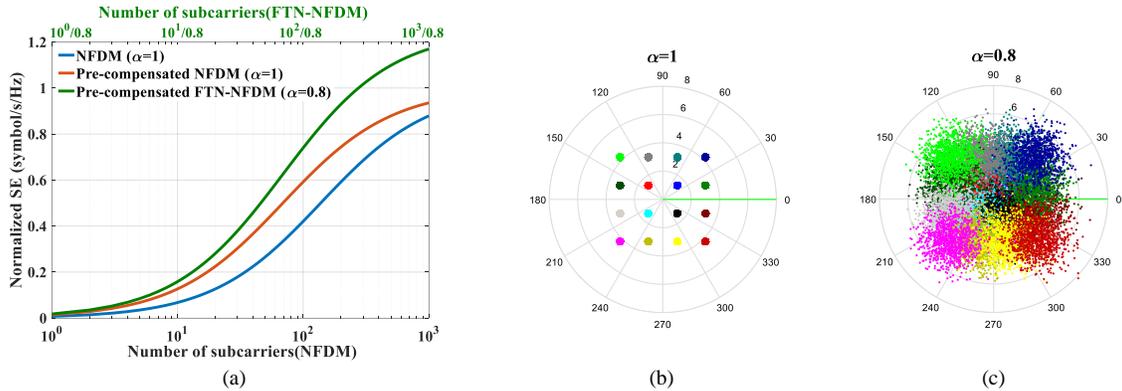

Fig.2 (a) The normalized SE of NFDM($\alpha$=1) and FTN-NFDM ($\alpha$=0.8) as function of the number of modulated nonlinear subcarriers, where the bandwidth is 32GHz and transmission distance is 1000km; (b) Demodulated QAM constellation in NFDM ($\alpha$=1). (c) Demodulated QAM constellation interfered by ICI in FTN-NFDM ($\alpha$=0.8).

## 4. Simulation results and discussion

The simulation platform is depicted in Fig.3. At the transmitter DSP, the QAM symbols are modulated on the $b(\lambda)$ with or without a compression factor. After PDC, the time domain signals are obtained by inverse NFT. The fiber channel comprises of 12 spans, and each span includes 80km SSMF (with attenuation $\alpha$=0.2dB/km, chromatic dispersion coefficient D=16.8ps/(nm× km), nonlinearity parameter $\gamma$=1.3/W/km), an EDFA with a 5dB noise figure and an optical band pass filter (OBPF) with 40GHz bandwidth. After coherent detection, the normalization and NFT operation is performed. Because the PDC is used, only half of the channel response is reversed (phase shift). For FTN-NFDM, the decoding is performed from the modulated coefficient $b(\lambda)$ and the iterative detection-sphere decoder is used to cancel the impact of ICI[13]. For NFDM, only OFDM decoding is conducted. Finally, the system performance is analyzed through Q factor directly from the BER.

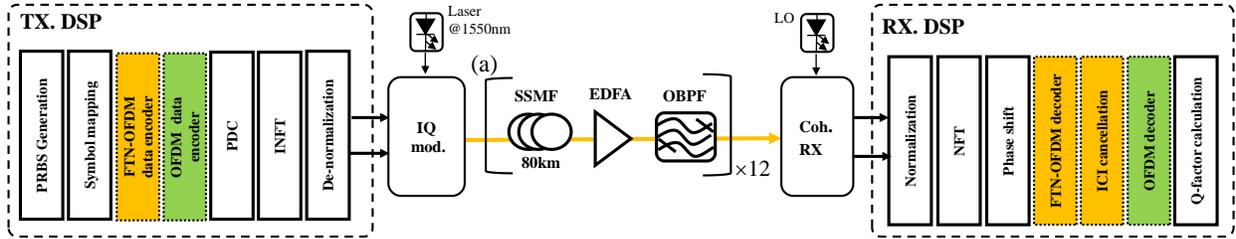

Fig.3. NFDM (FTN-NFDM) simulation platform. The yellow and green modules work only in the FTN-NFDM and NFDM system, respectively.

In our simulation, 32GHz signal bandwidth and 16 nonlinear subcarriers are used in NFDM system. For FTN NFDM system, 18 and 20 nonlinear subcarriers are used with the compression factor of 0.89 and 0.8, respectively. The modulation format is 16QAM. The total block time is 2.5ns and GI is 2ns. For NFDM, the normalized SE is 0.2(1/2.5ns×16/32GHz) and the net rate is 25.6Gbit/s (1/2.5ns×16×4bit/symbol). For FTN-NFDM system, the normalized SE is 0.225 and 0.25, and the net rate is 28.8Gbit/s and 32Gbit/s, respectively. The normalized SE is shown in Fig.4a. Compared to NFDM, the FTN-NFDM system increases SE by 12.5% (α=0.89) and 25% (α=0.8). When we use a smaller compression factor and a larger number of nonlinear subcarriers, the normalized SE will be increased and exceed 1, but the complexity of canceling ICI will be increased as well. Effective ICI canceling algorithms will be studied in our future work. We compared the performance of NFDM and FTN-NFDM transmission and the results are shown in Fig.4b. After 960km transmission, the FTN-NFDM system with the compression factor of 0.89 and 0.8 offer 0.7dB and 0.4dB performance improvement over NFDM system in terms of Q-factor, respectively.

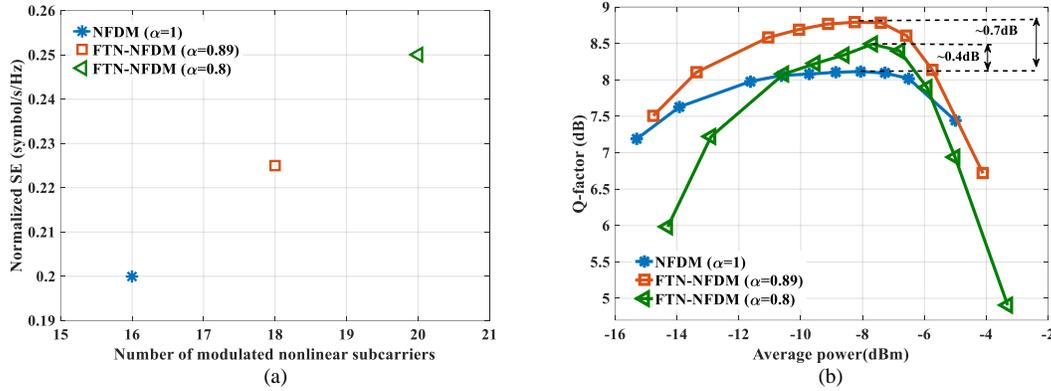

Fig.4. (a) The number of modulated nonlinear subcarriers and corresponding normalized SE (symbol/s/Hz) for NFDM system and FTN-NFDM system;(b) Q-factor as function of average power at the 960km fiber transmission for NFDM and FTN-NFDM system.

## 5. Conclusion

In this paper, to the best of our knowledge, we first proposed the FTN-NFDM system and the upper bound of FTN NFDM system was derived. Simulation systems with 960km transmitted distance and 16QAM were set up. Compared to NFDM system, FTN-NFDM system provided 12.5% and 25% increase in spectral efficiency and 0.4dB and 0.7dB Q-factor improvement through effective ICI cancelation algorithm for the compression factor of 0.89 and 0.8, respectively.